\documentclass[prd,aps,tightenlines,superscriptaddress,showpacs,nofootinbib,eqsecnum,amsfonts,amsmath,epsf]{revtex4}
\usepackage{bm,graphicx,epsfig}

\newcommand{\pa}{\partial}
\newcommand{\wt}{\hat{t}}
\newcommand{\wH}{\widehat{H}}
\newcommand{\ww}{\widehat{\omega}}
\newcommand{\wF}{\widehat{\cal F}}
\begin{document}
\title{  A Comparison of  search templates
 for gravitational waves from binary inspiral}

\author{Thibault Damour}

\affiliation{\it Institut des Hautes Etudes Scientifiques, 91440
Bures-sur-Yvette, France}

\author{Bala R. Iyer }
\affiliation{\it Raman Research Institute, Bangalore 560 080, India}

\author{B.S. Sathyaprakash}
\affiliation{\it School of Physics and Astronomy,
Cardiff University, 5, The Parade, Cardiff, CF24 3YB, U.K.}

\begin{abstract}
We compare the performances of the templates defined by three
different types of approaches: traditional post-Newtonian templates
 (Taylor-approximants), ``resummed'' post-Newtonian templates assuming 
 the adiabatic approximation and
stopping before the plunge (P-approximants),
 and further ``resummed'' post-Newtonian templates going beyond the adiabatic
approximation and incorporating the plunge with its transition from the
inspiral (Effective-one-body approximants). The signal to noise ratio
 is significantly enhanced  (mainly because of the inclusion of the
 plunge signal) by using these new effective-one-body templates relative to
 the usual post-Newtonian ones for a total binary mass 
 $m \agt 30 M_\odot$, and reaches a maximum around $ m \sim 80 M_\odot$. 
 Independently of the
question of the plunge signal, the comparison of the various templates
confirms the usefulness of using resummation methods.  The paper also
 summarizes the key elements of the construction of various templates 
 and thus can serve as a resource 
for those involved in writing inspiral search software.

\end{abstract}
\pacs{04.3.0Db, 04.25.Nx, 04.80.Nn, 95.55.Ym}
\date{\today}
\maketitle

\section {Introduction}

The late stage evolution in a compact binary, when the 
component stars are under the influence of the strong
gravitational fields of each other and are moving at 
relativistic speeds, is dictated by the non-linear dynamics 
of general relativity and is very difficult to model. 
In the early stages of adiabatic inspiral (that is, when the
inspiral time scale is much larger than the orbital time scale) it is
possible to treat the problem of motion perturbatively and to expand the 
general relativistic equations of motion and wave generation formulae in a 
power-series in $v/c,$ $v$ being a characteristic velocity.
(We henceforth use units such that $G = c = 1$.)
However, the phasing of the gravitational wave (GW) signal  derived from these
perturbative results becomes increasingly
 inadequate as the two bodies approach each other. 
The characteristic
 velocity  $v_p(m) \equiv ( \pi m f_p)^{1/3},$ corresponding to the peak of the
detector sensitivity to the inspiral signal from a binary of total mass 
$m = m_1 + m_2,$ is numerically equal (for initial LIGO; for which 
$f_p = 126$ Hz)\footnote{It should be noted that the LIGO noise curve used
in this paper is  the currently best available and
different from that used in DIS1 and DIS2.} 
to $v_p (m) = 0.125 (m/M_\odot)^{1/3}$. For a double 
neutron star system one already has $v_p(2.8) = 0.176$, while, for an 
archetypal  $ (10 M_\odot, 10 M_\odot)$ double black hole system 
one has $v_p(20) = 0.340$, quite close to the velocity corresponding 
to the last stable orbit (lso): $v_{\rm lso} \approx 1/\sqrt{6} = 0.408$. 

The present theoretical understanding has enabled the perturbative
 computation  (via post-Newtonian expansions) of the
binary orbit and GW phase to an order $v^5$ beyond the
standard quadrupole formula.
We shall use these $v^5$-accurate results in this work. 
At present, we cannot count on the (hopeful) extension of the 
post-Newtonian (PN) perturbative
calculations by another two orders, to order $v^7$, because currently used PN 
techniques leave undetermined a physically crucial parameter 
entering at the $v^6$ level \cite{js98,djs,bf00}.
Moreover, as emphasized some time ago\cite{cutleretal93a}, the PN series 
(which is essentially a Taylor expansion in powers of $v$)
 is a rather poorly convergent expansion.
More precisely, if one considers the PN expansion of the crucial GW flux 
(see, e.g., Fig.~3 of \cite{dis01}), one notices that the $v^4$-accurate and $v^5$-
accurate approximations start significantly deviating, in opposite directions,
from the exact (test-mass) result when $v\gtrsim 0.2$.
As such relatively high values of $v$ are typically involved in the calculation 
of the GW phasing ($v\sim v_p(m) \geq 0.18$ as soon as $m \geq 2.8 M_\odot$),
one has to worry that search templates based on a
straightforward use of PN-expanded results might be 
inadequate for the detection and/or measurement of inspiral signals, 
especially for the more massive systems 
($m\geq 10 M_\odot$ implies $v_p(m)\geq 0.34$) 
which are likely to be the first potentially detectable events.

 To address
this crucial problem, we have been advocating \cite{dis01,dis02,bd00}
a new philosophy for making the optimal use of existing PN results,
namely, to use several {\it re-summation techniques}
to improve the convergence of the  PN series, 
before using them to compute the GW phasing. 
As of now, we have proposed and studied 
three successive stages in the 
definition and use of such re-summation techniques. First, 
we constructed \cite{dis01} {\it time-domain} signals, 
called P-approximants --  starting from
the standard PN  Taylor representation -- which  possess better
convergence properties and capture the expected analytical behaviour
(poles and zeroes) of the relevant  physical quantities quite well. We have
shown that these new signal models, when compared with standard 
PN signals, are both more  {\it effectual } (larger overlaps) and 
 more {\it faithful} (smaller biases in the estimation of 
parameters) representations of some fiducial ``exact'' signals. 
Though {\it time-domain} P-approximants are better signal models
than the standard ``Taylor'' approximants,
they are  computationally  expensive to use in a data
analysis exercise that searches for inspiral signals using 
hundreds of thousands of templates, which have to be correlated, 
with arbitrary time lags, with the detector output. This data-analysis 
computational cost is much reduced (thanks to the existence of efficient
FFT algorithms) when one disposes of explicit analytic
expressions for the Fourier transform of the templates.

Second, we  found \cite{dis02}  explicit {\it frequency-domain}
representations of P-approximants (as well as of standard PN 
templates) that are {\it computationally inexpensive} and are yet 
as {\it faithful} and {\it effectual} as the original
 time-domain models. This frequency-domain representation incorporates the
``edge oscillations'' due to the (assumed) abrupt shut-off of the 
time-domain signal occurring soon after the binary crosses the last
stable orbit.
In \cite{dis02} we emphasized that the signal to noise ratio of the first 
interferometric detectors is large enough for detection only for massive
binary black hole systems of total mass $m\gtrsim 25 M_\odot$. For such 
systems the characteristic velocity corresponding to the peak of the 
detector sensitivity is $v_p(25)\simeq 0.37$, which is very close
to $v_{\rm lso}\simeq 1/\sqrt{6}$, the velocity at the last stable
orbit.
One, therefore, expects that 
the first detections are most likely to concern massive\footnote{ Note that,
even for less massive systems, the necessity to capture more than $96.5 \%$
of the SNR, corresponding to a loss in the number of events by no
more than $10 \%,$  implies that one must accurately control the phasing 
of the waveform at frequencies significantly higher than $f_p$, corresponding
to velocities significantly higher than $v_p (m)$.} systems 
($ 20  M_\odot  \lesssim m \lesssim 40  M_\odot)$
 with $v_p \sim v_{\rm lso}$.
 
It is therefore crucial to push the re-summation techniques 
introduced in \cite{dis01} further
 so as to be able to describe not only the GW phasing
 during the last cycles before lso crossing, but also during the 
transition\footnote{The P-approximants model this transition by 
a sharp cut-off in the signal.}
 between inspiral and plunge, and during the plunge itself. 
Recently,  Buonanno and Damour \cite{bd00}
 combined some of the re-summation techniques of \cite{dis01}
 and \cite{dis02}
with a novel approach to the general relativistic dynamics of two-body
systems\cite{bd99} to devise an improved type of re-summation approach to
the GW phasing of coalescing binaries, able to describe in 
more detail the transition between inspiral and plunge. This ``effective
one-body'' approach is the first one that goes beyond the ``adiabatic
approximation'', used both in standard (non-resummed) PN approximants,
and in P-approximants.

The data analysis of inspiral, merger and ring-down was pioneered
by Flanagan and Hughes \cite {fh98}. They treated the problem of
inspiral rather accurately but the 
merger/plunge was treated by assuming that about 10\%
of the rest mass energy would be emitted during merger.
This quite optimistic estimate was based on a crude model of the
coalescence of maximally spinning black holes, and was arbitrarily
extended to all cases. A similar back-of-the-envelope consideration
of the ring-down amplitude let them to optimistically assume that about 3\%
of the rest mass energy would be emitted during ring-down.

In this paper we discuss only non-spinning\footnote{In view of current
black hole binary formation mechanisms \cite{lpp97},
we think it likely that most
of them will include only slowly spinning holes.} binaries and we make
no ad hoc assumption about the total energy radiated during the merger
phase. The Effective One-Body (EOB) formalism does not treat the inspiral
and plunge phases separately. Indeed, in this formalism the plunge is seen
as a natural continuation of the inspiral phase contributing (for equal masses)
about 0.6 orbital cycles (or 1.2 GW cycles), with a total energy
associated with the plunge around 0.7\%. The energy emitted during the
following (matched) ring-down phase is also found to be around  0.7\% \cite
{BD-MG9}.
These energy losses are much
smaller than the Flanagan-Hughes guesstimates of 10\% and 3\%, respectively.
 Consequently,
it is unlikely that we will be able to detect the plunge
phase of the EOB waveforms separately, irrespective of the mass of the system.
This is in sharp contrast to the Flanagan-Hughes claim
that the SNR contribution of the sole merger phase of massive 
black holes of total mass in the
range 30-1000 $M_\odot$ will dominate over the inspiral phase contribution.
Note also that (for a source at 100 Mpc)
the (merger-dominated) SNR of \cite{fh98} reaches a maximum
of 40 around $m \sim 200 M_\odot$, while our (inspiral-dominated) SNR
reaches a maximum of 8 around $m \sim 80 M_\odot$.
 [It seems that most of the difference between Fig. 4 
of Flanagan and Hughes and our Fig.~\ref{fig:snr} below, e.g. a factor 3
(between their 25 and our 8) between the SNR for a 80 $M_\odot$ source
at 100 Mpc, comes from the huge difference in energy loss during merger.]

In this paper we make a prediction that the
merger phase will enhance the inspiral phase SNR
by about 10\% for $m \sim 30 M_\odot$, and 
by about 300\% for $m \sim 80 M_\odot$,
but that it will not be in itself a significant 
signal. As a result our best candidate sources are still
stellar mass black hole binaries of total mass in the range 
30-90 $M_\odot.$  We also conclude that 
the ring-down phase is in itself not a significant contribution
for $m \alt 200 M_\odot$. 

There is one word of caution regarding the plunge signal: Even though
the plunge lasts for only about half an orbital cycle its spectral content
spreads over a large frequency range.
Consequently, the number of frequency bins over which the signal
spreads out is quite large and it is not advisable to use a non-optimal
method to try to detect the plunge part in isolation.
In fact, we believe that one of the robust predictions of the EOB approach
(at least in the case of slowly spinning holes) is that the plunge signal
is a smooth continuation of the inspiral one, and that one should
 use templates that are phase-coherent all over the inspiral-plus-plunge
 phase. We are aware of the approximate nature of the EOB results (especially
 beyond the lso) and do not claim that the EOB waveform is the last word
 on the problem, but our position is the following: (i) in absence of 
 comparatively accurate alternative 
 results, it is important to study in detail the
 predictions coming from the EOB waveform, and (ii) we shall finally
 recommend to use a bank of filters which cover a large range of possibilities,
 with special weight being given to the best-tested ``resummed'' templates.
 
We also hope that our work will give an additional incentive to numerical
relativity groups toward computing waveforms which are at least as accurate
(and physically complete) as the EOB one. In particular, let us recall that
\cite{bd00} has proposed a new approach to the numerical computation of
binary black hole coalescences: namely, to start the numerical evolution
just after lso crossing, i.e. at a stage where one can still trust the
{\em resummed} PN estimate of the dynamics of two black holes, but where
there is only 0.6 orbit to evolve before coalescence. To this aim 
Ref.~\cite{bd00} has provided explicit results for the  initial dynamical
data (positions and momenta in ADM coordinates) of this problem. However,
apart from stimulating further thoughts on the problem (\cite{bbcl00}),
we are not aware of the existence of numerical simulations implementing
the proposal of Ref.~\cite{bd00}, nor are we aware of other numerical
work leading to explicit (non adiabatic) waveforms for coalescing binaries
which could be compared to the EOB one.

Data analysis groups associated with various ground-based interferometers
are now finalizing the  analysis software that will be used for
GW searches in data that is expected to become
available in a few years time. It is essential that these groups be
aware of recent theoretical developments and of
their respective merits so as to take the
best advantage of the current knowledge in writing their software.
With this view in mind 
the aim of this paper is two-fold:
First, we wish to compare the performances of the templates defined by the 
three different types of approaches mentioned above (traditional 
``non-resummed'' PN templates, ``resummed'' PN templates assuming the adiabatic approximation
and stopping before the plunge, and further ``resummed'' PN templates going
beyond the adiabatic approximation and
incorporating the plunge with its transition from the inspiral.)
Second, in view of the fact that the original 
publications \cite{dis01,dis02} are quite complex and technical,
we wish to  summarize in a more accessible manner
the key elements of the techniques introduced there (and re-used, with further
inputs, in \cite{bd00}.)
The present work should serve as an easily accessible resource 
for data analysis groups. Readers interested in a more detailed understanding
of our general approach are referred to \cite{dis01,dis02,bd00,bd99}
 for motivation, formalism, logical reasoning, exhaustive
tests and further  discussion of the new signal models.

\section {Time-domain phasing formulas in the adiabatic approximation}
\label{sec:time domain}

In searching for GW from an inspiralling compact binary we are
faced with the following data analysis problem: On the one hand, we have
some (unknown) exact gravitational waveform $h^X (t;\lambda_k)$ where 
$\lambda_k$,
$k=1, \ldots, n_{\lambda},$ are the parameters of the signal (e.g.,
the masses $m_1$ and $m_2$ of the members of the emitting binary).  
On the other hand, we have theoretical calculations of the motion 
of \cite{dd81,js98,djs,bf00}, and
gravitational radiation from \cite{bdiww95,b96}
binary systems consisting of neutron stars (NS) or black 
holes (BH) giving the PN expansions 
 of
an energy function $E(x\equiv v^2),$ which is related to the total relativistic 
energy $E_{\rm tot}$ via $E_{\rm tot}=(m_1+m_2)(1+E),$ and a GW
luminosity (or ``flux'') function ${\cal F}(v)$. Here, the dimensionless
argument $v \equiv x^{\frac{1}{2}}$ is an invariantly defined ``velocity''
related to the instantaneous  GW
frequency $F$ ($=$ twice the {\it orbital} frequency) by
$ v \equiv (\pi m F)^{\frac{1}{3}}$.
Given PN expansions of the motion of and gravitational radiation from
a binary system, one needs to compute
the {\it ``phasing formula''}, i.e. an accurate mathematical
model for the evolution of the GW
phase\footnote{ We work within the  ``restricted'' waveform
approximation which keeps only the leading harmonic in the GW signal.}
 $\phi^{\rm GW} = p[t;\lambda_i],$
involving the set of parameters $\{ \lambda_i \}$ carrying 
information about the emitting binary system.
In the adiabatic approximation
the phasing formula is easily derived from the energy and flux functions.
Indeed,
 the standard energy-balance equation $dE_{\rm tot} / dt =-{\cal F}$ gives the 
following parametric representation of the phasing formula:
\begin{equation}
t(v) = t_{\rm ref} + m \int_v^{v_{\rm ref}} dv \, 
\frac{E'(v)}{{\cal F}(v)}, \ \ 
\phi (v) = \phi_{\rm ref} + 2 \int_v^{v_{\rm ref}} dv v^3 \, 
\frac{E'(v)}{{\cal F}(v)},
\label {eq:phasing formula general1}
\end{equation}
where $t_{\rm ref}$ and $\phi_{\rm ref}$ are integration constants and
$v_{\rm ref}$ an arbitrary reference velocity. [It is sometimes
convenient, though by no means necessary, to take as $v_{\rm ref}$
the velocity $v$ at the last stable orbit (see below).] 
>From the view point of computational purposes it is more efficient
to work with the following pair of coupled, non-linear, ordinary differential
equations (ODE's) that are equivalent to the above parametric formulas:
\begin{equation}
\frac{d\phi}{dt} - \frac{2v^3}{m}=0, \ \ 
\frac{dv}{dt}  + \frac{{\cal F}(v)}{mE'(v)}=0.
\label {eq:phasing formula general2}
\end{equation}

We shall see later that, for massive systems, the adiabatic approximation fails and one must replace the  two ODE's by a more complicated ODE system.
We now turn to a discussion of what is known about 
the two functions $E(v)$ and ${\cal F}(v)$ entering the phasing 
formula and how that knowledge can be improved.

\subsection {$T$-approximants}

We denote by $E_{T_n}$ and ${\cal F}_{T_n}$ the $n^{\rm th}$-order\footnote{
The label $n$ always refer to an approximant accurate up to
$ v^n = x^{(n/2)}$ included.} 
``Taylor'' approximants (as defined by the PN expansion)
 of the energy and flux functions:
\begin {equation}
E_{T_{2n}}(x) \equiv E_N(x) \sum_{k=0}^n \hat{E}_k(\eta)x^k, \ \ \ \
{\cal F}_{T_n}(x) \equiv {\cal F}_N(x) \left [\sum_{k=0}^n \hat{{\cal 
F}}_k(\eta)v^k + 
\sum_{k=6}^n \hat{L}_k(\eta) \log(v/v_0) v^k \right],
\label{eq:energy1 and flux1}
\end {equation}
\begin{equation}
\mbox{where,}\ \  E_N(x) = -\frac{1}{2}\eta x,\ \ 
{\cal F}_N(x) = \frac{32}{5}\eta^2 x^{5}. 
\label{eq:newton}
\end{equation}
Here the subscript $N$ denotes the ``Newtonian value'',
$\eta\equiv m_1 m_2/m^2$ is the symmetric mass ratio, and
$v_0$ is a fiducial constant to be chosen below.
In the test mass limit, i.e. $\eta \rightarrow 0,$  $E(x)$
is known exactly, from which the Taylor expansion of 
$E_{T_n}(v,0)$ in Eq.~(\ref{eq:energy1 and flux1}),
can be computed to all orders. In the $\eta\rightarrow 0$
limit, the exact flux is known numerically
\cite {p95,tanakaetal96} and 
the Taylor expansion of flux in Eq.~(\ref{eq:energy1 and flux1}) 
is  known \cite {tanakaetal96,p93,cutleretal93b} up to order $n=11$. 
On the other hand, in the physically relevant case where $\eta$  is
 finite, the above Taylor approximants are 
known \cite {bdiww95,b96}  only up to five-halves PN order, i.e. $n=5.$
 Recently the energy function has been computed up to 3PN order {\it i.e.} $n=6$, though with the presence of an unknown parameter \cite{djs,bf00}.
All the completely known coefficients in the expansions are enlisted in Table \ref{table:energy function}.

The problem is to construct a sequence of approximate waveforms $h_n^A
(t;\lambda_k)$, starting from the PN expansions of $E(v)$
and ${\cal F}(v).$ In formal terms, any such construction defines a {\it map} 
from
the set of the Taylor coefficients of $E$ and ${\cal F}$ into the (functional) 
space
of waveforms. Up to now, the literature has considered (one of) the
 map(s), say $T$,
\begin {equation}
(E_{T_n}\,,\,{\cal F}_{T_n}) \stackrel{T}{\rightarrow} h_n^T (t,\lambda_k) \, , 
\end{equation}
obtained by inserting the successive Taylor approximants
into the phasing formula \cite {cutleretal93a,p95}.
For brevity, we often refer to these ``Taylor'' approximants as
`` T-approximants''.
It must also be emphasized that even within this Taylor family of templates,
there are   at least three ways of proceeding  further,
leading to the following three inequivalent constructs: 
\begin{enumerate}
\item[{t1.)}] One can retain the rational polynomial ${\cal F}_{T_n} / E_{T_n}$
as it appears in  Eq.~(\ref{eq:phasing formula general2}) 
and integrate the two ODE's numerically. We shall denote the phasing
formula so obtained as $\phi^{(1)}_{T_n}(t):$
\begin{equation}
\frac{d\phi^{(1)}}{dt} - \frac{2v^3}{m}=0, \ \ \ \
\frac{dv}{dt}  + \frac{{\cal F}_{T_n}(v)}{mE_{T_n}'(v)}=0.
\label {eq:phasing formula1}
\end{equation}
\item[{t2.)}] One can re-expand the rational function ${\cal F}_{T_n} / E_{T_n}$
appearing in the phasing formula and truncate it at order $v^n$,
 in which case the integrals
in Eq.~(\ref{eq:phasing formula general1}) can be worked out analytically, to obtain
a {\it parametric} representation of the phasing formula in terms of 
polynomial expressions in the auxiliary variable $v$ 
\begin {equation}
\phi^{(2)}_{T_n}(v) =  \phi^{(2)}_{\rm ref} +
\phi^v_N (v)\sum_{k=0}^{n} \hat{\phi}^v_k v^k, \ \ 
t^{(2)}_{T_n}(v) = t^{(2)}_{\rm ref} +t^v_N(v) \sum_{k=0}^{n} \hat{t}^v_k v^k, 
\label{eq:phasing formula2}
\end {equation}
where a superscript on
the coefficients (eg. $\phi_1^v$) indicates that $v$ is the expansion parameter 
[as is  explicit from Table~\ref{table:phasing formulas}, the coefficient of $\phi^v_k$ include
 in some cases,  a $\log v$ dependence].
\item[{t3.)}] Finally, the second of the polynomials in Eq.~(\ref{eq:phasing formula2}) can
be inverted and the resulting polynomial for $v$ in terms of
$t$ can be substituted in $\phi^{(2)}(v)$ to arrive at an explicit  time-domain
phasing formula
\begin{equation}
\phi^{(3)}_{T_n}(t) = \phi^{(3)}_{\rm ref}+\phi_N^t \sum_{k=0}^{n} 
\hat{\phi}^t_k\theta^k,\ \ \ \
F^{(3)}_{T_n}(t) =  F_N^t \sum_{k=0}^{n} \hat{F}^t_k \theta^k,
\label{eq:phasing formula3}
\end{equation}
where $\theta=[\eta (t_{\rm ref}-t)/(5m)]^{-1/8}$ and 
$F \equiv d \phi/ 2 \pi dt =v^3/(\pi m)$ is the instantaneous GW frequency.
The coefficients in these expansions are all listed in Table~\ref{table:phasing formulas}. 
\end {enumerate}

\subsection {$P$-approximants}

Before defining new ``resummed''
 energy and flux functions with improved performances
we digress for a brief reminder of Pad\'e re-summation,
 which is a standard mathematical
technique used to accelerate the convergence of 
poorly converging power series.
Let $S_n (v) = a_0 + a_1 \, v + \cdots + a_n \, v^n$ be a truncated Taylor
series.
A Pad\'e approximant of the function whose Taylor approximant to order $v^n$ is
$S_n$ is defined by two integers $m,k$ such that $m+k=n$. If $T_n [\cdots]$
denotes the operation of expanding a function in Taylor series and truncating it
to accuracy $v^n$ (included), the $P_k^m$ Pad\'e approximant of $S_n$ is
defined by
\begin{equation}
P_k^m (v) = \frac{N_m (v)}{D_k (v)}; \ \ T_n [P_m^k (v)] \equiv S_n (v),
\label{eq:A1}
\end{equation}
where $N_m$ and $D_k$ are {\it polynomials} in $v$ of order $m$ and $k$
respectively. If one assumes that $D_k (v)$ is normalised so that $D_k (0) =1$;
i.e. $D_k (v) = 1+q_1 \, v + \cdots$, one shows that Pad\'e approximants are
uniquely defined by Eq.~(\ref{eq:A1}). 
In many cases the most useful\footnote{
The rare theorems dealing with the
Pad\'e technique concern the convergence of ``near-diagonal'' Pad\'e
approximants, i.e.  $m \rightarrow \infty$ with $\vert m - k\vert $ fixed.}
 Pad\'e approximants are the
ones near the ``diagonal'', $m=k$, i.e. $P_m^m$ if $n=2m$ is even, and
$P_m^{m+1}$
or $P_{m+1}^m$ if $n=2m+1$ is odd. 
  In this work we shall use
the diagonal ($P_m^m$) and the ``sub-diagonal'' ($P^m_{m+1}$) approximants 
which can be conveniently\footnote{
A convenience of this form is that the $n$-th continued-fraction coefficient
$c_n$ (see below) depends only on the knowledge of the PN coefficients of order $\leq n$.}
written in a continued
fraction form \cite{bo84}. For example, given
$S_2 (v) = a_0 + a_1 \, v + a_2 \, v^2$
one looks for
\begin{equation}
P_1^1 (v) = \frac{c_0}{1+ \frac{c_1 v}{1+c_2 v}} = \frac{c_0 (1+c_2 v)}{1+(c_1
+c_2) v}. \label{eq:A3}
\end{equation}
The continued fraction Pad\'e coefficients of a series 
containing six terms, i.e. $S_5(v),$ are given by 
\begin{eqnarray}
c_0 & =  & a_0,\ \ 
c_1 = -\frac{a_1}{a_0},\ \ 
c_2 = -\frac{a_2}{a_1} + \frac{a_1}{a_0}, \ \ 
c_3 = \frac{a_0 (a_1 a_3 - a_2^2)}{a_1 (a_1^2 - a_2 a_0)} \nonumber \\
c_4 & = & -\frac{c_0 c_1 (  c_2 +  c_1 )^3+c_0 c_1 c_2 c_3 (c_3+2c_2+2c_1)-a_4 }
	   {c_0c_1 c_2c_3}, \nonumber \\
c_5 & = & - \frac{(( c_2+ c_1)^2 + c_2 c_3)^2}{c_2 c_3 c_4} 
          - \frac {(c_4 +c_3 +c_2 +c_1)^2}{c_4}  
          - \frac{a_5} {c_0 c_1 c_2 c_3c_4}\, . 
\label{eq:pade coefficients}
\end{eqnarray}

In \cite{dis01} and \cite{dis02} we introduced several techniques 
for ``re-summing'' the Taylor expansions (in powers of $v$) 
of the energy and flux functions.
Starting from the PN expansions of $E$ and $\cal F$, in DIS1
we proposed a new class
of waveforms, called {\it P-approximants,} 
based on two essential ingredients:
(i) the introduction, on theoretical grounds, of two new, supposedly 
more basic and hopefully better behaved, energy-type and flux-type 
functions, say $e(v)$ and $f(v)$, and (ii) the systematic use of 
Pad\'e approximants (instead of straightforward Taylor
expansions) when constructing successive approximants of the intermediate
functions $e(v)$, $f(v)$. Schematically, our procedure is based on the
following map, say ``P'':
\begin{equation}
(E_{T_n} , {\cal F}_{T_n}) \rightarrow (e_{T_n} , f_{T_n}) 
\rightarrow (e_{P_n}, f_{P_n}) \rightarrow (E[e_{P_n}] , 
{\cal F}[e_{P_n} , f_{P_n}])
\rightarrow h_n^P (t,\lambda_k). \label {eq:n5}
\end{equation}
Our new energy function $e(x),$ where $x\equiv v^2,$ 
is constructed out of the total relativistic energy 
$E_{\rm tot}(x)$ using
\begin{equation}
e(x) \equiv \left( \frac{E_{\rm tot}^2 - m_1^2
-m_2^2}{2m_1 m_2}\right)^2 -1. \label{eq:N9}
\end{equation}
The function $E(x)$ entering 
the phasing formulas is the total energy per unit mass after subtracting
out the rest mass energy: $E(x)=[E_{\rm tot}(x)-m]/m$ and
is given in terms of $e(x)$ by
\begin {equation}
E(x) = \left[1 + 2 \eta \left (\sqrt {1+e(x)} - 1 \right )\right]^{1/2}-1,
\ \ 
 \frac{dE}{dx} = \frac {\eta e'(x)}{2 \left [1+ E(x)\right] \sqrt {1+e(x)}}.
\label{eq:Eofx and Eprime}
\end{equation}
Note that the quantity $E'(v),$ needed in the phasing 
formula, is related to $dE(x)/dx$ via $E'(v)=2 v dE(x)/dx.$
In the test-mass limit $e(x)$ and $dE(x)/dx$ are known exactly:
\begin{equation}
e_{\eta=0}(x)=-x \frac{1-4x}{1-3x},\ \ 
E^{\eta=0}(x)=\eta\left(\frac{1-2x}{\sqrt{1-3x}}-1\right),\ \
 \frac{dE^{\eta=0}}{dx} = -\frac{\eta}{2}\frac{(1-6x)}{(1-3x)^{3/2}}.
\label {eq:ETMT}
\end{equation}
The rationale for using $e(x)$ as the basic quantity
rather than $E_{\rm tot}(x)$ can be found in \cite{dis01}; here 
we note the following two points: 
(1) In the test mass case $e(x)$ is meromorphic in the complex $x$-plane,
with a simple pole 
singularity, while the function $E(x)$ is non-meromorphic and
exhibits a branch 
cut. (2) Secondly, in the test mass case,
the Pad\'e approximant of $e_{T_{2n}}(x),$ for $n\ge 2,$ yields the known
exact expression including the location of the lso and the pole.
Therefore, the function $e(x)$ is more suitable in analyzing 
the analytic structure than is $E(x)$.
In the comparable mass case, under the assumption of structural
stability between the case $\eta \rightarrow 0$ and the case 
of finite $\eta,$ one can expect the exact
function $e(x)$ to admit a simple pole singularity on the real axis 
$\propto (x-x_{\rm pole})^{-1}.$ We do not know the
location of this singularity, but Pad\'e approximants
are excellent tools for giving accurate 
representations of functions having such pole singularities 
\cite {bo84}. 

Our proposal is the following: Given some usual Taylor
approximant to
the normal energy function, $E_{T_{2n}} = -\frac{1}{2} \, \eta \, x \,
(1+E_1 \, x
+ E_2 \, x^2 + \cdots + E_n \, x^n)$, one first computes the corresponding
Taylor approximant for the $e$ function, say
\begin {equation}
e_{{T}_{2n}}(x) =  -x \sum_{k=0}^n e_k x^k. 
\label{eq:23}
\end {equation}
(Remember that we 
consistently label the successive approximants by the order in 
velocity; e.g. a $2PN$-accurate object has the label 4.)
Then, one defines the Pad\'e approximant of 
$e_{T_{2n}}(x)$\footnote{ More precisely,
$e_{P_{2n}}(x)$ is $-x$ times the Pad\'e approximant of 
$-x^{-1} e_{T_{2n}}(x)$.}
\begin{equation}
e_{P_{2n}} \, (x) \equiv -x \, P_{m+\epsilon}^m \, 
\left[ \sum_{k=0}^{n} e_k \, 
x^k \right]
\label{eq:n25}
 \end{equation}
where $\epsilon =0$ or 1 depending on whether $n \equiv 2m+\epsilon$ is even or
odd. We shall call the continued fraction Pad\'e coefficients of $e_{P_{2n}}$
as $c_1, c_2, \cdots,$ (Note that $c_0\equiv 1$). 
They are given in terms of $e_k$  in Table \ref{table:energy function}.
Given a continued fraction approximant $e_{P_{2n}}(x)$ of the
truncated Taylor series $e_{T_{2n}}$ of the energy function $e(x)$
the corresponding $E_{P_{2n}}(x)$ and $dE_{P_{2n}}(x)/dx$ functions are 
obtained using formulas in Eq.~(\ref{eq:Eofx and Eprime}) by
replacing $e(x)$ on the right hand side with $e_{P_{2n}}(x).$

Apart from using it to improve the convergence of the PN series, 
Ref.~\cite{dis01} has also proposed to use the Pad\'e-resummed function
$e_{P_{2n}}(x)$ to determine the location of the lso, the Pad\'e
estimates of the lso being defined by considering the minima of $e_{P_{2n}}(x)$.
In contrast, in the Taylor case one must, for consistency, use
 the minima of $E_{T_n}(v)$  to define the 
locations of the lso. We have confirmed that in the test mass
case this Pad\'e-based method yields the exact result at orders
$v^4$ and beyond while the corresponding Taylor-based method [considering
the minima of $E_{T_n}(v)$] gives unacceptably high estimates
of $v_{\rm lso}$, i.e. of the GW frequency at the lso .
In the finite $\eta$ case, the  Pad\'e-resummed
predictions are in good qualitative, (and
reasonable quantitative) agreement with the more recent predictions based 
on the ``effective-one-body'' approach\cite{bd99}.
 The location of the lso's for the
various approximations together with the location of the light ring
[i.e. the pole singularity in $e_{P_{2n}}(x)$] are also tabulated in Table \ref{table:energy function}.

Having defined a new energy function, we move on to introduce a new
flux function. The aim is to define an analytic continuation
of the flux function to control its analytic structure as also to
handle the logarithmic terms that appear in the flux function in 
Eq.~(\ref{eq:energy1 and flux1}).
Factoring out the logarithmic terms is what  allows us to use 
standard Pad\'e techniques effectively in this problem.

It has been pointed out \cite{cutleretal93b} that the flux function 
in the test mass case ${\cal F}(v;\eta =0)$ has a simple pole at 
the light ring $v^2 = 1/3$. It has been argued that the origin of
this pole is quite general [cf. \cite{dis01}, discussion following Eq.~(4.3)]
and that even in the case of comparable masses we should expect to
have a pole singularity in ${\cal F}.$ However, as already pointed out, 
the light ring orbit in the $\eta\ne 0$ case 
corresponds to a simple pole $x_{\rm pole} (\eta)$ in the 
new energy function $e(x;\eta)$. Let us define 
the corresponding (invariant)
``velocity'' $v_{\rm pole} (\eta) \equiv \sqrt{x_{\rm pole} (\eta)}$. This
motivates the introduction of the following ``factored'' flux function,
$\hat{f}(v; \eta)$ 
\begin{equation}
\hat{f}(v;\eta) \equiv \left( 1-v/v_{\rm pole} \right) \, \hat{{\cal F}} (v;\eta).
\end{equation}
where $\hat{{\cal F}}(v)\equiv {\cal F}(v)/{\cal F}_N(v) = 5 {\cal F}(v)/(32 
\eta^2 
v^{10}),$ is the
Newton-normalised flux.  Note
that multiplying by $1-v/v_{\rm pole}$ rather than $1-(v/v_{\rm pole})^2$
has the
advantage of regularizing the structure of the Taylor series of $\hat{f}(v)$ in
introducing a term linear in $v,$ which is absent in the flux function in 
Eq.~(\ref{eq:energy1 and flux1}) (cf. Table \ref{table:phasing formulas}). Three
further inputs will allow us to construct better converging approximants to
$\hat{f}(v)$.  First, it is clear (if we think of $v$ as having the 
dimension of a velocity) that one should normalize the velocity $v$ 
entering the logarithms in the flux function in Eq.~(\ref{eq:energy1 and flux1})
to some relevant velocity scale $v_0$. In the  absence of further
information the choice $v_0 = v_{\rm lso} (\eta)$ seems justified (the other
basic choice $v_0 = v_{\rm pole}$ is numerically less desirable as $v$ will
never
exceed $v_{\rm lso}$ and we wish to minimize the effect of the logarithmic
terms).
A second idea, to reduce the problem to a series amenable to Pad\'eing, is to
factorize the logarithms. This is accomplished by writing 
the $\hat{f}$ function in the form
\begin{equation}
\hat{f}_{T_n}(v;\eta) = {\left( 1+ 
\sum_{k=6}^n \hat{l}_k v^k 
\ln\frac{v}{v_{\rm lso}} 
\right)} {\left( \sum_{k=0}^n \hat{f}_k v^k \right)}, 
\label{eq:n34}
\end{equation}
where coefficients $\hat{f}_k$ are 
$\hat{f}_0=1,$ 
$\hat{f}_{k+1}=\hat{{\cal F}}_{k+1} - \hat{{\cal F}}_k /v^{\rm pole}$
and $\hat{l}_k$ are constants determined from the coefficients of
$\hat{{\cal F}}_k$  by relations (analogous to) Eq. (4.9) of DIS1\footnote{The
variables 
$\hat{l}_k$ and $\hat{f}_k$ used  here are equal to the variables
$l_k$ and $f_k$ used in DIS1. They are  `hatted' here as a reminder
that they  represent coefficients of Newtonian-normalised
quantities.
The coefficients $A_k$ and $B_k$ appearing in the definition of $l_k$ are 
computed in Ref. \cite{tanakaetal96}.}.
Finally, we define Pad\'e approximants to the factored flux function 
$\hat{f}(v)$ as
\begin{equation}
\hat{f}_{P_n} (v;\eta) \equiv {\left( 1+ \sum_{k=6}^n \hat{l}_k v^k
\ln\frac{v}{v_{\rm lso}^{P_n} (\eta)} 
\right)} {P_{m+\epsilon}^m \left( \sum_{k=0}^n \hat{f}_k v^k \right)},
\label{eq:n35}
\end{equation}
where $v_{\rm lso}^{P_n} (\eta)$ denotes the lso velocity ($\equiv \sqrt{x_{\rm
lso}}$) for the $v^n$-accurate Pad\'e approximant of $e(x)$, and where
$P_{m+\epsilon}^m$ denotes as before a diagonal or sub-diagonal Pad\'e with $n
\equiv 2m+\epsilon$, $\epsilon =0$ or $1$. 
The corresponding approximant of the
flux $\hat{{\cal F}}(v)$ is then defined as
\begin{equation}
\hat{{\cal F}}_{P_n}(v;\eta) \equiv 
\left (1-\frac{v}{v_{\rm pole}^{P_n} (\eta)} 
\right)^{-1} \, \hat{f}_{P_n}(v;\eta), 
\label{eq:pade flux}
\end{equation}
where $v_{\rm pole}^{P_n}(\eta)$ denotes the pole velocity defined by the
$v^n$-Pad\'e of $e(x)$. In the test mass case the exact location of the
pole and the lso are $x_{\rm pole}=1/3$ and
$x_{\rm lso} = 1/6,$ respectively (cf. Table \ref{table:energy function}). We shall denote the continued
fraction Pad\'e coefficients of $\hat{f}_{P_n}(v)$ by $d_k$.
They can be found in terms of $\hat{f}_k$ using Eqs.~(\ref{eq:pade coefficients}).
At present, the most accurate estimate of the flux would be 
$\hat{{\cal F}}_{P_{11}}$, defined by using the known $\eta$-dependent 
coefficients $\hat{f}_k$ for $k \leq 5$, and the test-mass values of $\hat{f}_k$
and $\hat{l}_k$ for $k \geq 6$.

\section {Frequency-domain phasing formulas in the adiabatic approximation}
\label{sec:frequency domain}

The time-domain P-approximant waveforms discussed above  are 
computationally intensive to use in a full-scale data-analysis.
 Recently, in  DIS2 we have
constructed  frequency-domain representations of the $P$-approximants
which are 10-50 times faster to compute than their time-domain
analogues but are yet as accurate. This increases the usefulness of
 $P$-approximants in data analysis.

The Fourier representations are based primarily on 
a newly derived improved version of the 
stationary phase approximation appropriate to time-truncated signals.
 $P$-approximants
 cannot be modeled using the standard  stationary
phase approximation over the entire frequency domain. Indeed, close
to the last stable orbit, where the inspiral phase terminates,
 one requires a 
modification of the stationary phase approximation. In simple terms,
Ref.\cite{dis02} found a way of taking into account the effect of 
an assumed abrupt termination of the waveform near the last stable circular
orbit by introducing simple modifications to the usual stationary phase
approximation. We present only the final results here; the interested reader
is referred to DIS2 for details.
Note that the results summarized below are quite general and can 
be applied to a generic chirp signal which shuts off abruptly ({\it i.e.}
on a time scale $\lesssim F^{-1}$).

We begin with a discussion of the usual stationary phase approximation for
chirp signals.  Consider a signal of the form,
\begin {equation}
h(t)=2a(t)\cos\phi(t)= a(t) \left [ e^{-i \phi(t)} + e^{i \phi(t)} \right ],
\end {equation}
where $\phi(t)$ is the implicit solution of one of the phasing
formulas in Eq.~(\ref{eq:phasing formula1}), Eq.~(\ref{eq:phasing formula2})
or Eq.~(\ref{eq:phasing formula3}) 
for some choice of functions $E'$ and ${\cal F}$ \cite{code}. 

The quantity $2\pi F(t) = {d\phi(t)}/{dt}$ defines the instantaneous
GW frequency $F(t)$, and is assumed to be
continuously increasing. (We assume $F(t)>0$.)
 Now the Fourier transform $\tilde h(f)$ of $h(t)$ is defined as
\begin {equation}
\tilde{h}(f) \equiv \int_{-\infty}^{\infty} dt e^{2\pi ift} h(t)
            = \int_{-\infty}^{\infty}\,dt\, a(t) 
              \left[ e^{2\pi i f t - \phi(t)}  +  e^{2\pi ift +\phi(t)}\right ].
\end {equation}
The above transform can be computed in the stationary
phase approximation (SPA). For positive frequencies only the first term
on the right  contributes and yields the following {\it usual} SPA:
\begin {equation}
\tilde{h}^{\rm uspa}(f)= \frac {a(t_f)} {\sqrt {\dot{F}(t_f)}}
e^{ i\left[ \psi_f(t_f) -\pi/4\right]},\ \ 
\psi_f(t) \equiv  2 \pi f t -\phi(t), 
\label{eq:ft phase}
\end {equation}
and $t_f$ is the saddle point defined by solving for $t$, $ d \psi_f(t)/d t = 0$,
i.e. the time $t_f$ when the GW frequency $F(t)$ becomes equal to the 
Fourier variable $f$. In the (adiabatic) approximation where 
Eqs.~(\ref{eq:phasing formula general1}) hold, 
the value of $t_f$ is given by the following integral:
\begin{equation}
t_f = t_{\rm ref} + m \int_{v_f}^{v_{\rm ref}} \frac{E'(v)}{{\cal {\cal F}}(v)} 
dv,
\label{s1}
\end{equation}
where $v_f \equiv (\pi m f)^{1/3}.$
Using $t_f$ from the above equation and $\phi(t_f)$ in
Eq.~(\ref{eq:phasing formula general1}) one finds that
\begin{equation}
 \psi_f(t_f) = 2 \pi f t_{\rm ref} - \phi_{\rm ref} + 2\int_{v_f}^{v_{\rm ref}} 
(v_f^3 - v^3)
\frac{E'(v)}{{\cal {\cal F}}(v)} dv .\label{eq:s2}
\end{equation}
The big computational advantage of Eq.~(\ref{eq:s2}) [with respect to its
time-domain counterpart, Eq.~(\ref{eq:phasing formula general1})], is that,
 in the frequency domain, there 
are no equations to solve iteratively; the Fourier amplitudes 
are given as explicit functions of frequency.

In the Fourier domain too there are many inequivalent
 ways in which the phasing
$\psi_f$ can be worked out. Here we mention only the most popular:
\begin{enumerate}
\item[{f1.)}] Substitute (without doing any re-expansion or re-summation) 
for the energy and flux functions their PN expansions
or the P-approximants of energy and flux functions 
and solve the integral in Eq.~(\ref{eq:s2}) numerically
to obtain  the T-approximant SPA or P-approximant SPA, respectively.
\item[{f2.)}] Use PN expansions of energy and flux but
 re-expand the ratio $E'(v)/{\cal F}(v)$ in Eq.~(\ref{eq:s2}) in
which case the integral can be solved explicitly. This leads to the following
explicit, Taylor-like, Fourier domain phasing formula:
\begin{equation}
 \psi_f(t_f) = 2 \pi f t_{\rm ref} - \phi_{\rm ref} + 
 \tau_N \sum_{k=0}^5 \hat{\tau}_k (\pi m f)^{(k-5)/3} 
\label{eq:s3}
\end{equation}
where $\hat{\tau}_k$ are the chirp parameters listed in 
Table~\ref{table:phasing formulas}.
\end{enumerate}
Eq.~(\ref{eq:s3}) is 
one of the  standardly used  frequency-domain phasing formulas.
Therefore, we shall use that as one of the models in our comparison
of different inspiral model waveforms. We refer to it as ``type-f2''
frequency-domain phasing.

Just as in the time-domain, the frequency-domain phasing is 
most efficiently computed by a pair of coupled, non-linear, ODE's:
\begin{equation}
\frac{d\psi}{df} - 2\pi t = 0, \ \ \ \
\frac{dt}{df} + \frac{\pi m^2}{3v^2} \frac{E'(f)}{{\cal F}(f)} = 0,
\label {eq:frequency-domain ode}
\end{equation}
rather  than by numerically computing the integral in  
Eqs.~(\ref{eq:s2}).

Next we correct the performance of the usual SPA by including 
{\it edge corrections}
 arising as a consequence of modeling the time-domain
signal as being abruptly terminated at a time $t=t_{\rm max}$ 
({\it time-truncated chirp})
when the GW frequency reaches $F=F_{\rm max}.$  In practice, we expect that
$F_{\rm max}$ will be of order the GW frequency at the lso. However,
we prefer to leave unspecified
the exact value of $F_{\rm max}$. The idea is to use $F_{\rm max}$
as a free model parameter, to be varied so as to maximize the overlap
(see Section \ref{sec:conclusion})
 between
the $t_{\rm max}$-truncated template and the real signal.
 Such time-truncated signals can be represented as:
\begin {equation}
h(t)=2 a(t) \cos \phi (t) \theta ( t_{\rm max}  - t),
\end{equation}
where $\theta$ denotes the Heaviside step function, i.e. 
$\theta(x)=0$ if $x<0$ and $\theta(x)=1$ when $x\ge 0.$
The effect of this time-windowing has been modeled in DIS2 
and the result is that the Fourier transform of such a  time-truncated
signal can be accurately represented in the two regions 
$f \le F_{\rm max}$ and $f \ge F_{\rm max},$ by
\begin{eqnarray}
f\le F_{\rm max}& : & \tilde h_<^{\rm ispa}(f) = {\cal C}(\zeta_{<}(f)) 
\frac{a(t_f)}{\sqrt{\dot{F}(t_f)}} e^{i\left[\psi_f(t_f)-\pi/4\right]},
\nonumber \\
f \geq F_{\rm max} &:  &\tilde{h}_{>}^{\rm ispa}(f) = {\cal C}(\zeta_{>}(f))
\frac{a(t_{\rm max})}{\sqrt{\dot{F}(t_{\rm max})}} \exp{ \left [i\psi_f(t_{\rm max})+ 
i\frac{\pi (f-F_{\rm max})^2}{\dot{F}(t_{\rm max})}-i\pi/4\right]} \,,
\label{eq:3.37}
\end{eqnarray}
where the label `ispa' stands for improved SPA, ${\cal C}$ is 
essentially the complementary error
function, $2{\cal C}(\zeta) \equiv \rm erfc\left(e^{i\pi/4}\zeta\right),$
and $\zeta$ is computed using
\begin {eqnarray}
& & f<F_{\rm max}: \zeta_< \equiv 
- [{\psi_f(t_f)-\psi_f(t_{\rm max})}]^{1/2},\nonumber \\
& & f \ge F_{\rm max}: \zeta_>(f) = \frac{\sqrt{\pi}(f-F_{\rm max})}{\sqrt 
{\dot{F}(t_{\rm max})}}.
\label{eq:zeta}
\end {eqnarray}
The error function needed in calculating ${\cal C}(f)$  may be
 numerically computed using the NAG \cite{nag} library S15DDF.
 Note the denominator $\sqrt{\dot{F}(t_{\rm max})}$ entering 
Eqs.~(\ref{eq:3.37}) and (\ref{eq:zeta}).
 We define a generic time-truncated signal as a chirp for which this
 denominator is finite (neither infinite, nor zero).
  [These signals were called ``Newtonian-like''in Ref.~\cite{dis02}.
  We do not keep this name here to emphasize that the results 
Eqs.~(\ref{eq:3.37}) and (\ref{eq:zeta}) 
  apply also to relativistic models, as long as $\dot{F}(t_{\rm max})$ is finite.]

 The exceptional (non-generic) case where $\sqrt{\dot{F}(t_{\rm max})}$ becomes
 infinite
arises if one tries to keep using the simple adiabatic phasing approximation
up to the last stable orbit defined by the corresponding (approximate) 
energy function $E(v)$. This exceptional case can also be dealt with at the
price of a more complicated modification of the usual stationary phase
result (see Ref. \cite{dis02} for details).  We do not enter into details
here because the recent work \cite{bd00} on the transition
between the adiabatic inspiral and the plunge has shown that the adiabatic 
approximation breaks down just before the lso, and that $\dot{F}(t)$ never
becomes infinite. 

\section{Time-domain phasing formula beyond the adiabatic approximation}
\label{sec:plunge}

In the above, we restricted ourselves to the standard 
``adiabatic approximation'', where one estimates the phasing by combining the energy-balance equation $dE_{\rm tot}/dt=-{\cal F}$ with some resummed
estimates for the energy and flux as  functions of the instantaneous circular
orbital frequency. Recently, Buonnano and Damour \cite{bd00} have introduced
a new approach to GW from coalescing binaries which is no longer limited
to the adiabatic approximation, and which is expected to describe 
rather accurately the transition between the inspiral and the plunge, and to give also a first estimate of the following plunge signal. The approach
of \cite{bd00} is essentially, like \cite{dis01,dis02}, a re-summation
technique which consists of two main ingredients: (i) the ``conservative''
(damping-free) part of the dynamics (effectively equivalent to the
specification of the  $E(v)$ in the previous approaches) 
is resummed by a new technique which replaces the
two-body dynamics by an effective one-body dynamics \cite{bd00}, and
(ii) the ``damping'' part of the dynamics [equivalent to the specification of
the  ${\cal F}(v)$]
is constructed by borrowing the re-summation technique introduced in 
\cite{dis01}. In practical terms, the time-domain waveform 
is obtained as the following function of the reduced time $\hat{t}=t/m$:
\begin{equation}
\label{4.1}
 \quad h(\wt) = {\cal
C} \, v_{\omega}^2 (\wt) \cos (\phi_{\rm GW} (\wt))\,, \quad v_{\omega}
\equiv \left( \frac{d \varphi}{d \wt} \right)^{\frac{1}{3}}\,, \quad \phi_{\rm
GW} \equiv 2\varphi \,.
\end{equation}
The orbital phase $\varphi(\wt)$ entering Eq.~(\ref{4.1}) is given by
integrating a system of ODE's:
\begin{mathletters}
\label{eq:4.2}
\begin{eqnarray}
\label{eq:3.28}
&&\frac{dr}{d \wt} = \frac{\pa \wH}{\pa p_r}
(r,p_r,p_\varphi)\,, \\
\label{3.29}
&& \frac{d \varphi}{d \wt} = \ww \equiv \frac{\pa \wH}{\pa p_\varphi}
(r,p_r,p_\varphi)\,, \\
\label{3.30}
&& \frac{d p_r}{d \wt} + \frac{\pa \wH}{\pa r}
(r,p_r,p_\varphi)=0\,, \\
&& \frac{d p_\varphi}{d \wt} = \wF_\varphi(\ww (r,p_r,p_{\varphi}))\,.
\label{3.31}
\end{eqnarray}
\end{mathletters}
The reduced Hamiltonian $\widehat{H}$ (of the associated one-body problem) 
is given, at the 2PN approximation, by 
\begin{mathletters}
\label{eq:4.4}
\begin{eqnarray}
\label{eq:3.32}
\wH(r,p_r,p_\varphi) = \frac{1}{\eta}\,\sqrt{1 + 2\eta\,\left [
\sqrt{A(r)\,\left (1 + \frac{p_r^2}{B(r)} + \frac{p_\varphi^2}{r^2} \right )} -1 \right ]}\,,\\
\label{3.34}
{\rm where}\;\;
A(r) \equiv 1 - \frac{2}{r} + \frac{2\eta}{r^3} \,,
\quad \quad B(r) \equiv \frac{1}{A(r)}\,\left (1 - \frac{6\eta}{r^2}
\right )\,.
\end{eqnarray}
\end{mathletters}
The 3PN version of $\widehat{H}$ 
has been recently obtained \cite{djs}.
The damping force $\cal{F}_{\varphi}$ is approximated by
\begin{equation}
\widehat{{\cal F}}_{\varphi}
=-\frac{1}{\eta v_\omega ^3}{\cal F}_{P_n}(v_\omega)\,,
\label{eq:damp}
\end{equation}
where $
{\cal F}_{P_n} (v_{\omega})  
= \frac{32}{5}\,\eta^2\,v_\omega^{10}\,
\hat{{\cal F}}_{P_n} (v_{\omega})$  
is the flux function used in P-approximants discussed above.

The system  Eq.~(\ref{eq:4.2})
 allows one to describe the smooth transition which takes place 
between the inspiral and the plunge (while the system 
(\ref{eq:phasing formula general2})
becomes spuriously singular at the lso, 
where $E'(v_{\rm lso})=0$). Ref.\cite{bd00} advocated to
continue using Eqs.~(\ref{eq:4.2}) after the transition, 
to describe the waveform emitted during the plunge
and to match around the ``light ring'' to a ``merger'' 
waveform, described, in the first approximation, 
by the ringing of the least-damped quasi-normal mode
of a Kerr black hole.[see Eq.~(6.2) of \cite{bd00}].
This technique is the most complete which is available at present.
It includes (in the best available 
approximation and for non-spinning black holes) most of the 
correct physics of the problem, and leads to a specific prediction for the
complete waveform (inspiral + plunge + merger) emitted by coalescing binaries.
Because of its completeness, we shall use it as our ``fiducial exact''
waveform in our comparison between different search templates.

The initial data needed in computing this effective one-body waveform are as follows:
In gravitational wave data analysis we are normally given an initial frequency
$f_0$ ($\hat \omega_0 \equiv \pi m f_0$) corresponding to the lower cutoff of a detector's sensitivity
window, at which to begin the waveform. 
The initial phase of the signal will not be known in advance but in order
to gauge the optimal performance of our approximate templates we maximise
the overlap (see Section \ref{sec:conclusion})
over the initial phases of both the fiducial exact
signal (i.e., the effective-one-body waveform) and the approximate template.
The general analytical result of this maximisation was discussed in
Appendix B of DIS1. In the terminology of this Appendix, these fully
 {\em phase-maximised} overlaps were called 
the {\it best} overlaps (they are given by Eq. (B.11) of DIS1).
As discussed in Appendix B of DIS1, there are two distinct measures
of the closeness of two signals: the {\em best} overlap (maximised over the
phases of both the template and the exact signal), and the {\em minimax} 
overlap (maximised over the template phase, with the worst possible exact phase).
In an investigation such as ours (where we are interested in the optimal
 mathematical closeness between different signals), the {\em best}
overlap is the mathematically cleanest measure of closeness of two families of
templates, and we shall use it here.
In addition, we shall also maximise over the other template parameters
(in particular, the masses) to get an intrinsic measure of the closeness
of two families of templates. Note that the resulting {\em fully maximised}
overlaps are different from the {\it maximised ambiguity function}
of Ref. \cite{sathyaprakash and dhurandhar} and the {\it fitting factor} of
Ref. \cite{apostolatos}. The latter (identical, but given different names
by different authors) quantities are well-defined measures of the closeness
of two signals {\em only} within the (simplifying) 
approximation where signals in
quadrature are orthogonal. This is, however, not the case for the signals
we consider, and at the accuracy at which we are working.

  For the computation of the best overlaps, it is sufficient to construct 
two signal waveforms, and two template waveforms, one with phase
equal to 0 and another with phase equal to $\pi/2.$
 The rest of 
the initial data $(r_0, p_{r}^0, p_{\varphi}^0)$ are found using
\begin{equation}
r_0^3 \left [ \frac {1 + 2 \eta (\sqrt{z(r_0)} -1 )}{1- 3\eta/r_0^2} \right ]-  \hat \omega_0^{-2} = 0,\ \ 
p^0_\varphi = \left [\frac {r_0^2 - 3 \eta}{r_0^3 - 3 r_0^2 + 5 \eta} \right ]^{1/2}r_0,\ \ 
p^0_r = \frac {{\cal F}_\varphi(\hat \omega)}{C(r_0,p^0_\varphi) (dp^0_\varphi/dr_0)}\ \ 
\end{equation}
where $z(r)$ and $C(r,p_\varphi)$ are given by
\begin{equation}
z(r) = \frac{r^3 A^2(r)}{r^3-3r^2+5 \eta},\ \ 
C(r,p_\varphi) = \frac{1}{\eta \wH (r,0,p_\varphi)
 \sqrt{z(r)}} \frac{A^2(r)}{(1-6\eta/r^2)}.
\end{equation}
The plunge waveform is terminated when the radial coordinate attains the value
at the light ring $r_{\rm lr}$ given by the solution to the equation,
\begin{equation}
r_{\rm lr}^3 - 3 r_{\rm lr}^2 + 5 \eta = 0.
\end{equation}
The subsequent ``merger'' waveform is constructed as in Ref.\cite{bd00}.

\section{Results and Conclusions}
\label{sec:conclusion}

In this section we compare the performances of various signal models 
by  choosing as fiducial  ``exact'' signal model 
the effective one-body waveforms discussed in the previous Section. 
An important yardstick for comparing different waveforms is
the {\it overlap}: Given two waveforms $h$ and $g$ their overlap 
is defined as 
\begin{equation}
{\cal O} (h,g) = \frac {\left < h, g \right >}
{\left <h, h\right >^{1/2}\left <g, g\right >^{1/2}}.
\end{equation}
In the above equation the scalar product $\left<,\right>$ is defined as
\begin{equation}
\left <h, g\right > = 2 \int_0^\infty \frac{df}{S_h(f)} \tilde h(f) \tilde g^*(f) + C.C.
\label{overlap}
\end{equation}
where C.C. denotes complex conjugation and $S_h(f)$ is the one-sided detector 
noise spectral density 
($S_h^{\rm one-sided}=2 S_h^{\rm two-sided}$
leading to  the factor 2 in Eq.~(\ref{overlap}), 
compared to the definition used in \cite{dis02},  where we always use 
the two-sided noise).
See the Appendix below for the noise performances of the various detectors.

Firstly, in Fig.~\ref{fig:snr} we compare the signal-to-noise ratios (SNRs),
expected in GEO, LIGO and VIRGO, 
for equal mass binaries located at 100 Mpc,
when detecting an ``exact'' signal $h$ by means of a bank of templates $k$:
\begin{equation}
\rho\equiv \frac{S}{\overline{N}}=\frac{\vert 
\langle k, h\rangle \vert}{\langle k,k\rangle^{1/2}}
=\vert {\cal O}(k,h)\vert\langle h,h\rangle^{1/2}.
\label{d6}
\end{equation} 
Thick lines plot the SNR obtained when $ k = h$, i.e. when the template
 perfectly matches our
fiducial exact waveform (i.e. the effective one-body 
waveform including its ``ringing tail''), 
and thin lines show how that gets degraded when
we use for $k$ the best post-Newtonian template $T^{f2}$ 
[cf. Eq.~(\ref{eq:s3})] truncated at the test-mass lso, 
$F^{\rm GW}_{\rm lso} =4400 M_\odot/m $ Hz, assuming still that
the true signal $h$ is the one-body effective waveform. 
(As usual, see, e.g., Section IV A of \cite{dis02}, we averaged over
all the angles.)
The overlaps ${\cal O}(k,h)$
are maximised over the time lag and the two phases (as explained in the
previous section), as well as over the two individual masses $m_1$ and $m_2$.\footnote
{The plots are jagged because we have 
computed the SNR numerically
by first generating the waveform in the time-domain and then
using its discrete Fourier transform in  Eq.(\ref{d6})}.
The greater SNR achieved by effective one-body 
waveforms for higher masses, as compared to Fig 1 of DIS2, is 
due to the plunge phase present in these 
waveforms. We have checked that the final merger signal, modelled as a
quasi-normal-mode, has a numerically insignificant effect in both the 
SNR and the overlaps: the 
overlap between our fiducial exact waveform and an effective 
one-body waveform minus quasi-normal modes is greater than 
0.98 for double black holes of masses smaller than $(40,40)M_\odot$.

The origin of the enhancement in SNR is easily understood.
The plunge waveform begins around $f \simeq F_{\rm lso}$ and lasts until
 $f \simeq 2 F_{\rm lso}$ \cite{bd00}. 
The mass of a binary whose last stable orbit velocity 
is equal to the characteristic velocity $v_p$ of LIGO's peak sensitivity is 
$m=34.8 M_\odot$.
Therefore, effective one-body waveforms from binaries of masses in the 
range $35 M_\odot \alt m \alt 70 M_\odot$
 have larger SNRs for LIGO, than the usual SPA models.
It is, therefore, crucial to go beyond the waveforms given in
the adiabatic approximation to take advantage of these higher
SNRs for larger masses. Indeed, the SNRs being as high as 8,
for one detector and a source at $100$ Mpc,
a network of four detectors (two LIGOs, GEO and VIRGO) 
will be able to reliably search for such systems as
far as 150-200 Mpc.

 Next, in Table \ref{table:overlaps} we show the {\it fully maximised}
overlaps of effective one-body waveforms with 
signal models $T^{\rm t1}$ (column 2), $T^{\rm t3}$ (column 3), $T^{\rm f1}$ (column 4), 
$T^{\rm f2}$ (column 5) and $P$ (column 6) for four typical binaries.
The time-domain $T^{\rm t3}$-approximants are terminated when
$\dot{F}=0$ and the other  approximants
 are terminated when $F(t)=f_{\rm lso},$  
the lso being determined  consistently using $ E'_A(v)=0$, 
where  $E_A(v)$ is the corresponding approximate energy function. 
The overlaps are computed with the (initial) LIGO noise curve given
in the appendix.

  In addition to maximization  over the time lags and the phases 
as explained earlier,
 we also
 maximize over the masses of the approximate waveforms keeping the masses of the exact waveforms
fixed.  We note that none of the models have good overlaps with the ``exact''
one for
heavier mass binaries. This is as expected since it is for heavier masses that the
characteristic plunge phase makes a significant
 difference between the approximate and
the exact models.  The relative performances of
the  2PN and 2.5PN  Taylor-templates depends on the choice  of the scheme
used as is evident from columns three and four in the Table.
This is consistent with the results of Ref. \cite{tichy}. (The numerical
results must not be compared since they quote values for
the ``advanced'' LIGO.) Some Taylor models are ``effectual''
(large maximized overlaps), but at the cost of high biases in the parameters
 (i.e, in the terminology of \cite{dis01} they are not ``faithful'').
For example in the case of a fiducial exact (1.4,10) $M_\odot$ system
$T^{t1}_5$ reaches 0.9452  for mass values (0.8,20) $M_\odot$! Thus 
T-approximants  in the time-domain are significantly inferior to the other
models, both for their erratic convergence properties ( the $v^5$-accurate
templates being worse than the $v^4$-accurate ones) and for their poor
parameter estimation performance.

To further explore the performance of the various models we plot
in Fig.~\ref{fig:fdot} their intrinsic frequency evolutions in the LIGO band,
i.e. we plot $ \dot{F}/F^2$ versus $F$. The plot 
corresponds, in the fiducial exact case,  to a binary black hole of 
(10,10) $M_\odot$.
For the approximate models we exhibit the frequency evolution of
the system that achieves the maximum overlap. As expected, the maximum
overlap is obtained for template parameters such that the intrinsic frequency
evolution of the template waveform is ``tangent'' to the exact one, near
 the maximum sensitivity of the detector. This can always be achieved by
 fitting the mass parameters. The question is whether such a local ``tangency''
 ensures a sufficiently good ``global'' agreement.
 For instance, we note that the $T^{\rm t1,3}$
2.5 PN models fare poorly in  {\it globally} 
mimicking the frequency evolution of the
exact waveform. This is consistent with their returning
 the worst overlaps of all. On the other hand, even though the
$T^{f2}$ models do not reproduce the exact model over as large a range as
the $P$-approximants, they achieve nearly as large overlaps as the 
$P$-approximants, because they can be made (by optimizing the masses) to agree
well with the exact model over most of the sensitive part of the LIGO band.
 The $P$-approximants
are able to mimic the ``exact'' evolution the best with little bias in the masses
but, being based on the adiabatic approximation, they fail to capture the
smooth transition to plunge\footnote{We have confirmed that the 
$P$-approximants return the best overlaps when truncated at the
$P-$defined lso. Maximizing over a cutoff frequency smaller than the 
$P-$defined $F_{\rm lso}$ (which turns out to be higher than the 
effective-one-body-defined one) 
does not improve the overlaps.}.
The filters using the effective one-body approach go 
beyond the adiabatic approximation and include a smooth 
transition to plunge and merger. They, therefore,
supersede the adiabatic-limited P-approximants. This difference
between the two re-summed versions of binary signal models is important
for masses larger than about 20 $M_\odot.$

Note that, following \cite{bd00}, we have generated the effective-one-body
model using the adimensional time $\hat t=t/m$. [This trivially extends 
beyond the 2PN approximation \cite{djs}.] It has been recently
emphasized \cite{BSS00} (in the context of the  $T$- and $P$-approximants,
where one can also simplify formulae by working with $\hat t=t/m$)
that there are many computational advantages in working with such
adimensionalized time models. Indeed,  the phase evolution becomes 
completely independent of the total mass of the system. 
This, together with the fact that the
evolution can be computed from a system of ODE's, makes the computation
and storage of templates required in a search for binary black holes and neutron
stars in interferometer data computationally inexpensive as compared to
the conventional method that uses a 2-dimensional lattice of templates.

We are currently estimating the 
effects of unknown parameters in the 3PN motion and wave generation
\cite{DIJS00}.
The extension of the type of work presented here
 to go beyond the restricted post-Newtonian
approach, and also to include the effects due to spin and eccentricity, needs
to be systematically investigated.

To conclude: we believe that many of the new technical tools developed in
\cite{dis01,dis02} and briefly summarized above are useful ingredients 
for constructing effectual and fast-computed inspiral templates. 
For example: (i) the specific Pad\'e-based resummation
of the GW flux introduced in \cite{dis01}
is an important ingredient of the construction
 in \cite{bd00} of an accurate non-adiabatic waveform and, (ii) the improved 
 SPA technique derived in \cite{dis02} could be used to derive analytical
 approximations to the frequency-domain version of these effective-one-body
 waveforms. 
In view of our ignorance of the ``exact'' waveform emitted near and 
after the lso crossing, the best strategy is probably to construct a bank 
of templates which cover a large range of possibilities with special 
weight being given to the templates incorporating the best tested re-summation 
methods (such as P-approximants, and the effective-one-body-approach\cite{bd00}).
Because of the admittedly quantitatively rough, but plausibly qualitatively
correct, description of the plunge signal given by the EOB approach, we also
recommend to include some sort of multi-parameter template which qualitatively looks
like Fig. 12 of \cite{bd00}, but which introduces some flexibility both in the
phasing evolution during the plunge, and in the location of the matching to the
 ring-down (with the possible inclusion of several quasi-normal modes).
Finally, we emphasize the importance of modelling the transition to the plunge, 
and of including the signal emitted during the plunge: this leads to a
very significant enhancement of the signal-to-noise ratio, from about $4.5$
to $8$, for a source at 100 Mpc.

\acknowledgements

The GEO noise curve is based on data provided by G. Cagnoli and J. Hough,
LIGO is a fit to the envelope of the data from K. Blackburn, 
TAMA is from M.-K. Fujimoto and VIRGO is from J-Y Vinet.
We thank Alessandra Buonnano for  sharing her Mathematica codes
for waveform generation with us.
BRI and BSS would like to thank IHES, France and 
AEI, Germany for  hospitality during the final stages of this work.

\appendix

\section{ Noise Power Spectrum of Initial Interferometers}
In this short appendix we list the expected one-sided noise power spectral densities
of the various ground-based interferometers (Table \ref{table:noise psd}) and plot
the effective noise \footnote{Our effective noise $h_n$ is the same as what
has been conventionally called $h_{\rm rms}$ in the literature.}
$h_n\equiv \sqrt{fS_h(f)}$ in Fig.~\ref{fig:noise psd}.

\begin {table}[b]
\caption {Taylor coefficients of the energy functions
$E_{T_n}(x)$ and $e_{T_n}(x)$ and the corresponding
 location of the lso and pole.
As there are no terms of order $v^{2k+1}$ we have exceptionally chosen 
(for this Table only) the expansion 
parameter to be $x\equiv v^2$ rather than $v.$ In all cases the
 $k=0$ coefficient
is equal to 1, the last stable orbit is defined only for $k\ge 1$ in the
case of T-approximants and for $k\ge 2$ in the case of P-approximants
and $N$ denotes the ``Newtonian value''.}
\begin {center}
\begin {tabular}{c c c c}
$k$    & $N$ & 1  & 2 \\[3pt]
\hline\\[-8pt]
$\widehat{E}_k$ 
     & $-\frac{\eta v^2}{2}$
     & $-\frac{9+\eta}{12}$
     & $-\frac{81-57\eta+\eta^2}{24}$\\[3pt]
$e_k$  
      & $-x(=-v^2)$
      & $-\frac{3+\eta}{3}$
      & $-\frac{36-35\eta}{12}$\\[3pt]
$e_{P_k}$  
      & $-x(=-v^2)$
      & $c_1=\frac{3+\eta}{3}$
      & $c_2=-\frac{144-81\eta+4 \eta^2}{36+12\eta}$\\[3pt]
$x^{\rm lso}_{T_k}$
      & ---
      & $\frac{6}{9+\eta}$
      & $\frac {-E_1 + (E_1^2-3E_2)^{1/2}}{3E_2} $\\[3pt]
$x^{\rm lso}_{P_k}$
      & ---
      & ---
      & $\frac{-1 + (-c_1/c_2)^{1/2}}{c_1+c_2}$ \\[3pt]
$x^{\rm pole}_{P_k}$
      & ---
      & ---
      & $\frac{4(3+\eta)}{36-35\eta}$\\[3pt]
\end {tabular}
\end {center}
\label{table:energy function}
\end {table}

\begin {table}
\caption {Taylor coefficients of the flux, phase, time and frequency.
$N$ denotes the ``Newtonian value'' and 
$\theta=[\eta (t_{\rm lso}-t)/(5m)]^{-1/8}.$ In all cases the $k=0$ coefficient
is 1 and the  $k=1$ coefficient is zero. In certain cases
the 2.5 PN term involves $v^5 \log v$ or $\theta^5 \log \theta$
 term rather than a  $v^5$ or $\theta^5$ term.  
In those cases  we conventionally include the $\log v$ dependence in the
listed coefficient.
Chirp parameters $\tau_k,$ $k\ge 1,$
are related to the expansion parameters $t^v_k$ and $\phi^v_k$ via
$\tau_k = ( 8 \phi^v_k - 5 t^v_k )/3.$ 
We have given the simplified 
expressions for these in all cases, except $k=5$ where no simplification occurs
due to the presence of the log term in $\phi^v_5.$}
\begin {center}
\begin {tabular}{cccccc}
$k$ &  $N$ & 2  &  3  &  4  &  5  \\
\hline\\[-8pt]
$\widehat{{\cal F}}_k$ 
      & $\frac{32\eta^2 v^{10}}{5}$
      & $- \frac{1247}{336} - \frac{35\eta}{12}$
    & $4\pi$
    & $-\frac{44711}{9072} + \frac{9271\eta}{504} + \frac{65\eta^2}{18}$
    & $-\left(\frac{8191}{672} + \frac{583\eta}{24}\right) \pi$\\[3pt]
$\hat{t}^v_k$
      & $-\frac{5m}{256 \eta v^8}$
      & $\frac{743}{252} + \frac{11\eta}{3}$
      & $-\frac{32\pi}{5}$
      & $\frac{3058673}{508032} + \frac{5429\eta}{504} + \frac{617\eta^2}{72}$
      & $-\left(\frac{7729}{252}- \frac{13}{3}\eta\right)\pi$\\[3pt]
$\hat{\phi}^v_k$
      & $-\frac{1}{16\eta v^5}$
      & $\frac{3715}{1008}+\frac{55\eta}{12}$
      & $-10 \pi$
      & $\frac{15293365}{1016064} + \frac{27145\eta}{1008 } + 
\frac{3085\eta^2}{144}$
      & $ \left (\frac{38645}{672} - \frac{65\eta}{8 } \right ) \pi 
\ln \left ( \frac{v}{v_{\rm lso}} \right ) $\\[3pt]
$\hat{\phi}^t_k$
      & $-\frac{2}{\eta \theta^5}$
      & $\frac{3715}{8064}+\frac{55\eta}{96}$
      & $-\frac{3\pi}{4}$ 
      & $\frac{9275495}{14450688}+\frac{284875\eta}{258048 } + 
\frac{1855\eta^2}{2048 }$
      & $\left (\frac {38645}{21504} - \frac{65\eta}{256 } \right ) \pi 
\ln \left ( \frac {\theta}{\theta_{\rm lso}} \right ) $\\[3pt]
$\widehat{F}^t_k$
      & $\frac{\theta^3}{8\pi m}$
      & $\frac{743}{2688}+\frac{11\eta}{32}$ 
      & $-\frac{3\pi}{10}$
       & $\frac {1855099}{14450688} + \frac{56975\eta}{258048 } + 
\frac{371\eta^2}{2048 }$ 
      & $- \left(\frac{7729}{21504} - \frac{13}{256}\eta\right)\pi$\\[3pt]
$\hat{\tau}_k$ 
      & $\frac{3}{128\eta}$
      & $ \frac{5}{9} 
        \left ( \frac{743}{84} + 11\eta\right )$
      & $-16\pi $
      & $2  \phi^v_4$
      & $ \frac{1}{3} \left ( 8 \phi^v_5 - 5 t^v_5 \right ) $
\end {tabular}
\end {center}
\label{table:phasing formulas}
\end {table}

\begin {table}
\caption{Fully maximised overlaps of the fiducial exact ($X$) waveform 
(effective-one-body signal 
[7]) 
with: (1) the standard time-domain 
post-Newtonian approximations of type t1 and t3 ($T^{\rm t1}$, $T^{\rm t3}$), 
given in Eq.~(\protect{\ref{eq:phasing formula1}}) and 
Eq.~(\protect{\ref{eq:phasing formula3}}) and 
Table \protect{\ref{table:phasing formulas}}, (2) the
frequency-domain usual stationary phase approximations of type
1 and 2, ($T^{\rm f1}$, $T^{\rm f2}$) 
given in Eqs.~(\protect{\ref{eq:s2}}), (\protect{\ref{eq:ft phase}}) 
and (\protect{\ref{eq:s3}}) 
and Table \protect{\ref{table:phasing formulas}} and
(3) the time-domain P-approximants ($P$) -- energy function as given by 
Eqs.~(\protect{\ref{eq:Eofx and Eprime}}) and Eqs.~(\protect{\ref{eq:23}}),
flux function in Eqs.~(\protect{\ref{eq:n35}}) 
and (\protect{\ref{eq:pade flux}}), and coefficients enlisted in
Tables \protect{\ref{table:energy function}} 
and \protect{\ref{table:phasing formulas}}.
The overlaps, which are computed using the LIGO noise curve,
are maximised not only over the time-lag and 
 the initial phases of both the fiducial exact
signal and the approximate template
(by using two signal and two template waveforms, with phases
equal to 0 and $\pi/2$ 
[5]),
but also  over the two masses $m_1$ and $m_2$ of the approximate signal models. 
(The optimal masses are given below the overlaps.)
The time-domain $T^{\rm t3}$-approximants are terminated when
$\dot{F}=0$ and the other signals  are terminated when $F(t)=f_{\rm lso},$  
the lso frequency being determined consistently using $ E'_A(v)=0$
where $ E_A(v)$ is the corresponding approximate energy function.}
\begin{tabular}{cccccc}
$ k $ & $\left < X,T^{\rm t1}_k \right >$ 
      & $\left < X,T^{\rm t3}_k \right >$  
      & $ \left <X,T^{\rm f1}_k \right >$  
      & $ \left <X,T^{\rm f2}_k \right >$  
      & $ \left < X,P_k \right > $ \\
\hline
& \multicolumn{5}{c}{$m_1=m_2=15 M_\odot$}\\
\hline
4 & 0.8881 & 0.9488 & 0.8644 & 0.8144 & 0.8928\\
& (15.2,14.1) & (16.3, 16.4) & (14.7, 14.9) & (16.0,16.1) & (14.7, 15.1)\\
5 & 0.8794 & 0.8479 & 0.7808 & 0.8602 & 0.8929\\
& (17.3, 16.4) & (17.6, 17.9) & (16.8, 16.7) & (15.2,14.4) & (15.4,14.3)\\
\hline
& \multicolumn{5}{c}{$m_1=m_2=10M_\odot$}\\
\hline
4 & 0.9604 & 0.9298 & 0.9581 & 0.9109 & 0.9616\\
& (10.1,9.6) & (10.5, 10.3) & (10.0, 9.7) & (10.5, 10.6) & (10.0, 10.2)\\
5 & 0.8814 & 0.8490 & 0.8616 & 0.9529 & 0.9610\\
& (11.4, 10.6) & (11.4, 11.7) & ( 10.7, 11.0) & (10.3, 9.7) & (10.4, 9.7)\\
\hline
& \multicolumn{5}{c}{$m_1=10M_\odot, m_2=1.4M_\odot$} \\
\hline
4 & 0.9847 & 0.9673 & 0.9835 & 0.9721 & 0.9937\\
& (1.27,11.1) & (0.95,16.6) & (1.27, 11.1) & (0.96, 16.4) & (1.35, 10.5)\\
5 & 0.9452 & 0.6811 & 0.9394 & 0.9922 & 0.9941\\
& (0.82, 20.4) & (1.11, 15.7) & (0.82, 20.4) & (1.34, 10.5) & (1.37, 10.2)\\
\hline
& \multicolumn{5}{c}{$m_1=m_2=1.4M_\odot$} \\
\hline
4 & 0.8828 & 0.8538 & 0.8830 & 0.8503 & 0.9719\\
& (1.40, 1.39) & (1.42, 1.39) & (1.40, 1.39) & (1.44, 1.37) & (1.47, 1.34)\\
5 & 0.8522 & 0.7643 & 0.8522 & 0.9994 & 0.9945\\
& (1.46, 1.35) & (1.43, 1.38) & (1.46, 1.35) & ( 1.45, 1.35) & (1.49, 1.32)
\label{table:overlaps}
\end{tabular}
\end {table}

\begin {table}[h]
\caption {One-sided noise power spectral densities (PSD)
 of initial interferometers,
$S_h(f)$. For each detector the noise PSD is given in
terms of a dimensionless frequency $x=f/f_0$ and rises steeply above a lower
cutoff $f_s.$ }
\centering
\begin {tabular}{cccc}
Detector & $f_s$/Hz & $f_0$/Hz & $10^{46} \times S_h(x)$/Hz$^{-1}$ \\
\hline
GEO      & 40       &  150     & $\left [(3.4 x)^{-30} + 34 x^{-1} + 20(1-x^2+ x^4/2)/(1+x^2/2) \right ]$  \\
LIGO-I   & 40       &  150     & $9.00 \left [ (4.49x)^{-56} + 0.16 x^{-4.52} + 0.52 + 0.32 x^2\right ]$\\
TAMA     & 75       &  400     & $75.0 \left [x^{-5} + 13x^{-1}+9(1+x^2)\right ] $ \\
VIRGO    & 20       &  500     & $3.24 \left [ (6.23 x)^{-5} + 2 x^{-1} + 1 + x^ 2 \right]$\\
\end {tabular}
\label{table:noise psd}
\end {table}

\begin{figure}
\centerline {\epsfxsize 6 true in  \epsfbox {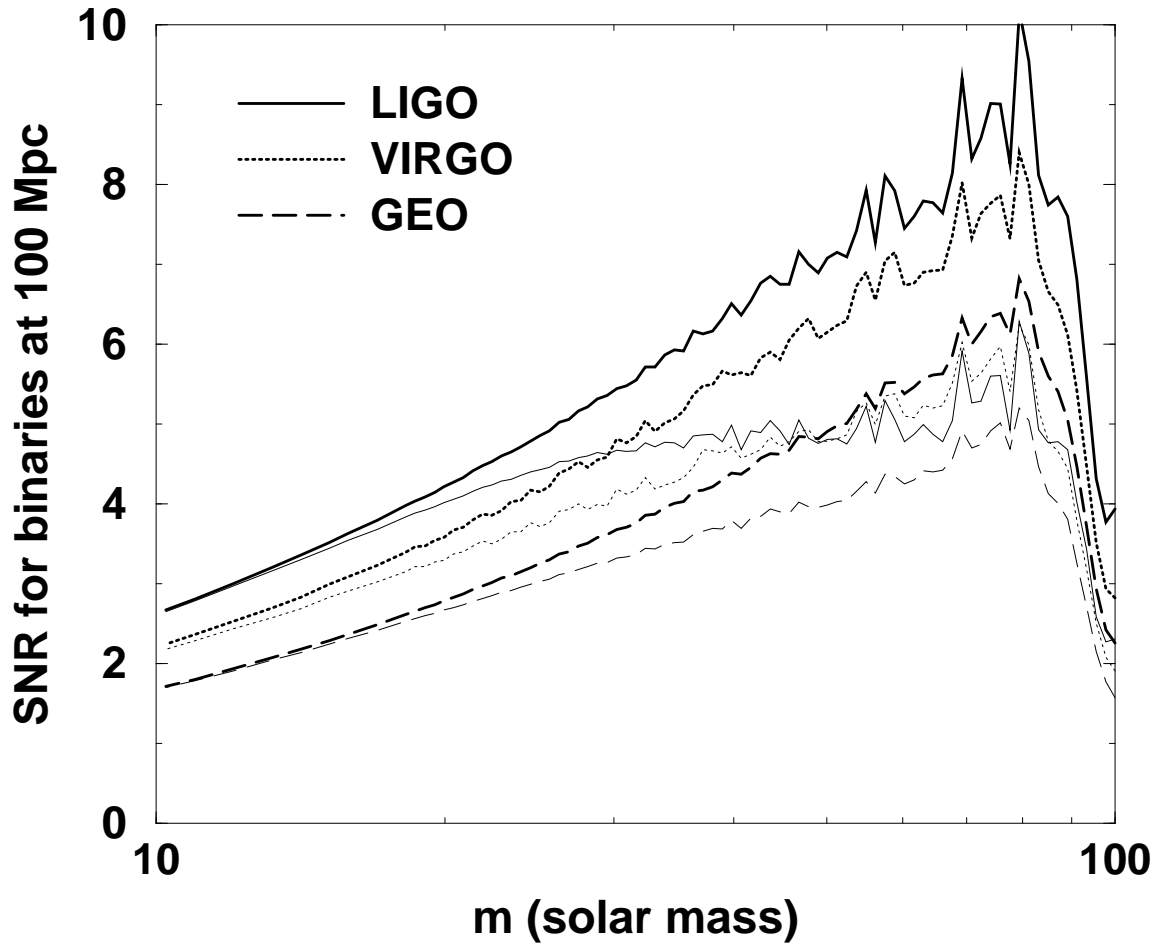} }
\caption{Signal-to-noise ratios in GEO, LIGO-I and VIRGO 
when using as Fourier-domain template the post-Newtonian model Eq.~(\ref{eq:s3})
($T^{f2}$),
truncated at the test-mass $F_{\rm lso}= 4400 M_\odot/m$ Hz (in thin lines),
compared to the optimal one obtained when the template coincides with the
fiducial ``exact'' (effective one-body) signal (thick lines).  
As usual, we averaged over
all the angles.
The overlaps are maximised over the time lags, the phases, and the
 two individual masses $m_1$ and $m_2$.
The plots are jagged because we have 
computed the SNR numerically
by first generating the fiducial ``exact'' waveform in the time-domain and then
using its discrete Fourier transform in  Eq.(\ref{d6}).
The greater SNR achieved by effective one-body 
waveforms for higher masses, as compared to Fig 1 of DIS2, is 
due to the plunge phase present in these waveforms.
Observe that the presence of the
plunge phase in the latter significantly (up to a factor of 1.5) 
increases the SNR for masses $m>35M_\odot.$ Using the effective one-body 
templates will, therefore, enhance the search volume of the
interferometric network by a factor of 3 or 4.}
\label{fig:snr}
\end{figure}

\begin{figure}
\centerline {\epsfxsize 6 true in  \epsfbox {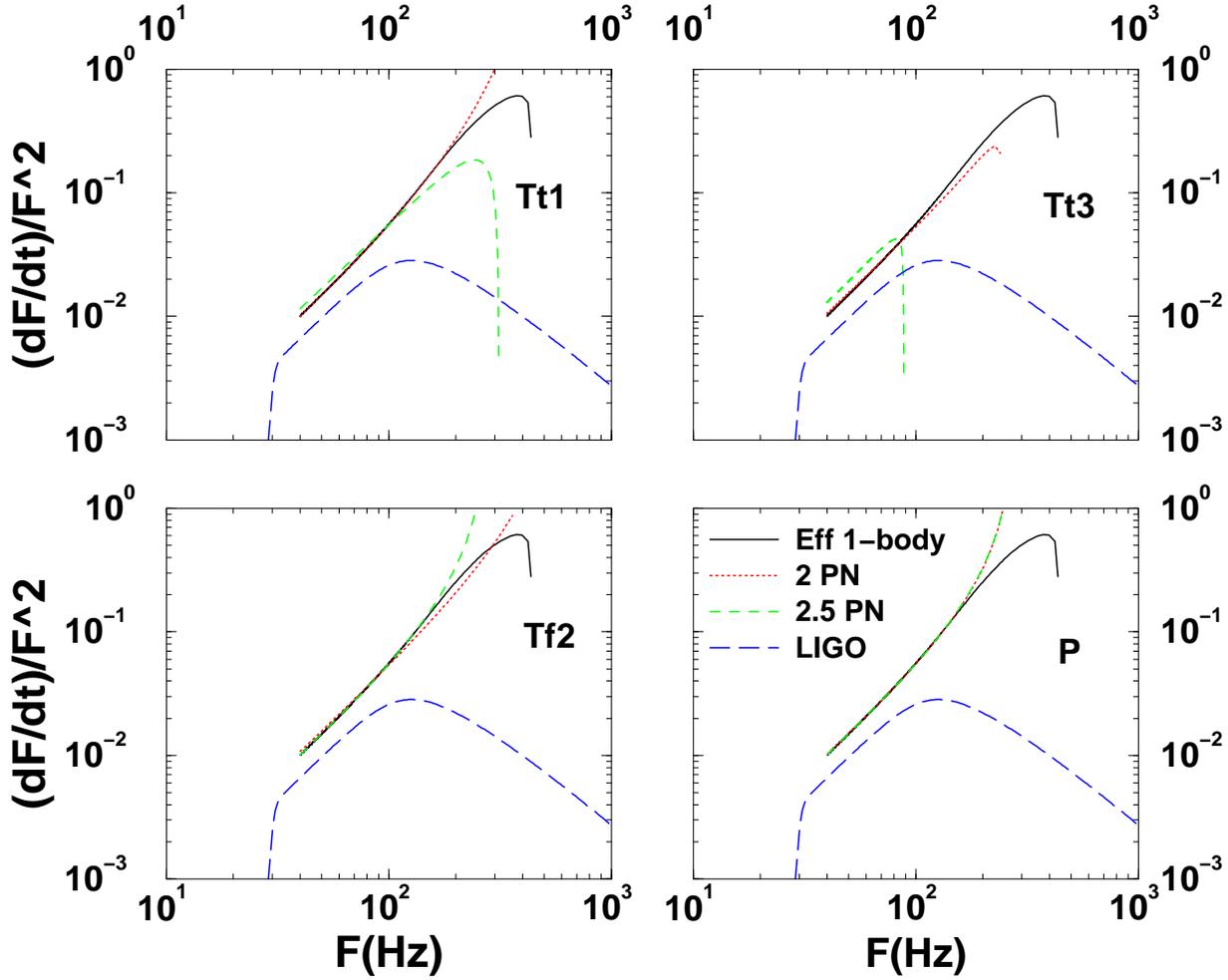} }
\caption{The frequency evolution of the various approximate
models is compared with the fiducial exact $(10,10) M_\odot$ model in the
LIGO band. 
To indicate the effect on the overlap, we also plot the
weighting function $1/h_n(f)$ for initial LIGO (not to scale),
which is a measure of the detector's sensitivity.}
\label{fig:fdot}
\end{figure}

\begin{figure}
\centerline {\epsfxsize 6 true in  \epsfbox {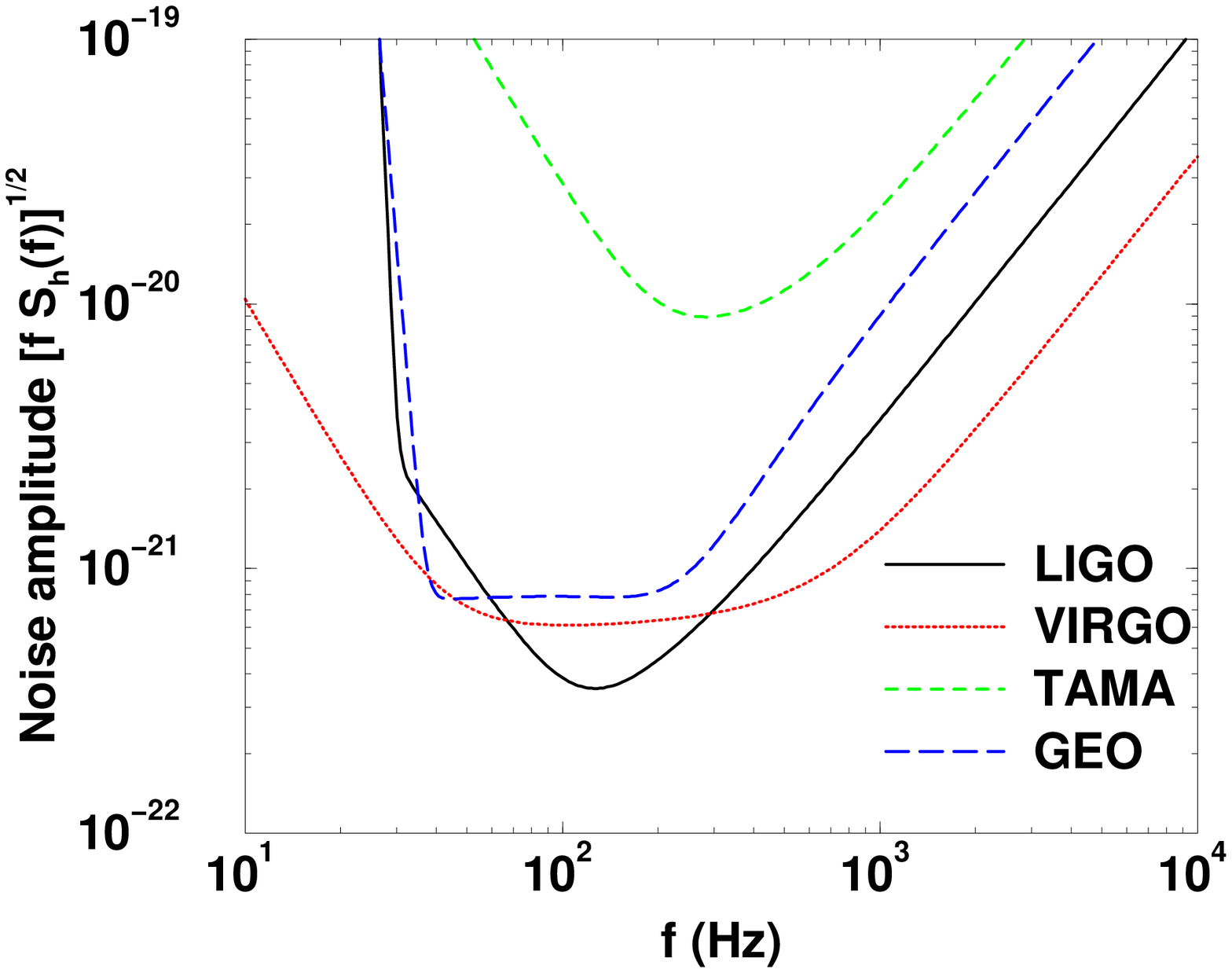} }
\caption{The effective noise $h_n=\sqrt{fS_h(f)}$ in various 
ground-based interferometers.}
\label{fig:noise psd}
\end{figure}


\begin{references}
\bibitem{js98}
P. Jaranowski and G. Sch\"afer, Phys. Rev. D {\bf 57}, 5948 (1998);
{\it ibid} D {\bf 57}, 7274 (1998); {\it ibid} D {\bf 60}, 1240003 (1999).
\bibitem{djs}
T. Damour, P. Jaranowski and G. Sch\"afer,
 Phys. Rev. D {\bf 62}, 021501; {\it ibid} D {\bf 62}, 044024; {\it ibid} D
{\bf 62}, 084011 (2000).
\bibitem{bf00}
L. Blanchet and G. Faye, Phys. Lett., {\bf 271,} 58 (2000); L. Blanchet and 
G. Faye, gr-qc/0004009; gr-qc/0006100; gr-qc/7051.
\bibitem{cutleretal93a}
 C. Cutler et al., Phys. Rev. Lett. {\bf 70}, 2984 (1993).
\bibitem{dis01}
T. Damour, B.R. Iyer, and B.S. Sathyaprakash,
Phys. Rev. D {\bf 57}, 885 (1998); referred to as DIS1
in this paper.
\bibitem{dis02}
T. Damour, B.R. Iyer, and B.S. Sathyaprakash,
Phys. Rev. D {\bf 62}, 084036  (2000);
referred to as DIS2 in this paper.
\bibitem{bd00} A. Buonanno and T. Damour,
 Phys.Rev. D {\bf 62}, 064015 (2000). 
\bibitem{bd99} A. Buonanno and T. Damour,
 Phys.Rev. D  {\bf 59}, 084006 (1999).
\bibitem{fh98}
E.E. Flanagan and S.A. Hughes, Phys.Rev. {\bf D57,} 
4535 (1998); {\it ibid} D57 (1998) 4566-4587.
\bibitem{lpp97}
V.M. Lipunov, K.A. Postnov, and M.E. prokhorov, New Astron. {\bf 2,} 43 (1997);
S.P. Zwart and S. McMillan, {\it Black hole mergers in the Universe,}
gr-qc/9910061.
\bibitem {BD-MG9} A. Buonanno and T. Damour, {\it Binary black holes
coalescence: transition from adiabatic inspiral to plunge,} contributed
paper to to the IX Marcel Grossmann Meeting in Rome, July 2000, gr-qc/0011052.
\bibitem {bbcl00} J. Baker, B. Brugmann, M. Campanelli and C. O. Lousto,
Class. Quant. Grav. {\bf 17,} L149 (2000)
\bibitem{dd81} T. Damour and N. Deruelle, Phys. Lett. {\bf 87A}, 81 (1981);
and C.R. Acad. Sci. Paris {\bf 293} (II) 537 (1981);
T. Damour, C.R. Acad. Sci.  Paris
{\bf 294} (II) 1355 (1982); and in {\it Gravitational Radiation},
ed. N. Deruelle and T. Piran, pp 59-144 (North-Holland, Amsterdam, 1983).
\bibitem{bdiww95} L. Blanchet, T. Damour, B.R. Iyer, C.M. Will and A.G.
Wiseman, Phys. Rev. Lett. {\bf 74}, 3515 (1995);
 L. Blanchet, T. Damour and B.R. Iyer, Phys. Rev. D {\bf 51},
5360 (1995);
 C.M. Will and A.G. Wiseman, Phys. Rev. D {\bf 54}, 4813 (1996);
 L. Blanchet,  B.R. Iyer, C.M. Will and A.G.
Wiseman, Class. Quantum. Gr. {\bf 13}, 575, (1996).
\bibitem {b96}
L. Blanchet, Phys. Rev. D {\bf 54}, 1417 (1996).
\bibitem{p95} E. Poisson, Phys. Rev. D {\bf 52}, 5719 (1995).
\bibitem{tanakaetal96} T.Tanaka, H.Tagoshi and M.Sasaki, Prog. Theor. Phys.
{\bf 96,} 1087  (1996).
\bibitem{p93} E. Poisson, Phys. Rev. D {\bf 47}, 1497 (1993).
\bibitem{cutleretal93b} C. Cutler, L.S. Finn, E. Poisson and
G.J. Sussmann, Phys. Rev. D {\bf 47}, 1511 (1993).
\bibitem{bo84} C.M. Bender and S.A. Orszag, {\it Advanced mathematical
methods for scientists and engineers} (McGraw Hill, Singapore, 1984).
\bibitem{code} At the Newtonian order, in terms of chirp mass,
${\cal M}=\eta^{3/5}m$, the  chirp
 amplitude
$a(t)=   {\cal C}_{{\cal M}} ( \pi {\cal M} F(t))^{2/3}$,
phase
$\phi(t) = \phi_c - 2\,
\left[ {(t_c-t)}/{5 {\cal M}} \right]^{5/8}$,
and frequency of the gravitational waves
$\pi {\cal M} F(t) = \left[ {5 {\cal M}}/{256(t_c-t)} \right]^{3/8}$.
\bibitem{nag} The Numerical Algorithms Group Limited, Oxford, United
Kingdom.
\bibitem {sathyaprakash and dhurandhar} B.S. Sathyaprakash and 
S.V. Dhurandhar, Phys. Rev. D {\bf 44,} 3819 (1991).
\bibitem {apostolatos} T. Apostolatos, Phys. Rev. {\bf D52,} 605 (1995).
\bibitem{tichy}
 W. Tichy, E. E. Flanagan, E. Poisson,
Phys.Rev. D {\bf 61}, 104015 (2000).
\bibitem{BSS00}
B.S. Sathyaprakash, {\it Mother Templates for computing gravitational wave chirps,} Class. Quant. Grav. (in press), gr-qc/0010044.
\bibitem{DIJS00}
T. Damour, B.R. Iyer, P. Jaranowski and B.S. Sathyaprakash, work in progress.
\end{references}
\end {document}